\documentclass[fleqn,10pt]{wlscirep}
\usepackage[utf8]{inputenc}
\usepackage[T1]{fontenc}
\usepackage{bm}

\title{Efficient and accurate neural field reconstruction using resistive memory}

\author[1,2,3]{Yifei Yu}
\author[1,2,3]{Shaocong Wang}
\author[4,5,6]{Woyu Zhang}
\author[1,2,3]{Xinyuan Zhang}
\author[1]{Xiuzhe Wu}
\author[1,2,3]{Yangu He}
\author[1,2,3]{Jichang Yang}
\author[1,2,3]{Yue Zhang}
\author[1,2,3]{Ning Lin}
\author[1,2,3]{Bo Wang}
\author[1,2,3]{Xi Chen}
\author[1,2,3]{Songqi Wang}
\author[7]{Xumeng Zhang}
\author[1]{Xiaojuan Qi}
\author[1,2,3,*]{Zhongrui Wang}
\author[4,5,6,*]{Dashan Shang}
\author[4,5,7*]{Qi Liu}
\author[2,8]{Kwang-Ting Cheng}
\author[4,5,7]{Ming Liu}

\affil[1]{Department of Electrical and Electronic Engineering, the University of Hong Kong, Hong Kong, China}
\affil[2]{ACCESS – AI Chip Center for Emerging Smart Systems, InnoHK Centers, Hong Kong Science Park, Hong Kong, China}
\affil[3]{Institute of Mind, the University of Hong Kong, Hong Kong, China}
\affil[4]{Key Lab of Fabrication Technologies for Integrated Circuits, Institute of Microelectronics, Chinese Academy of Sciences, Beijing 100029, China}
\affil[5]{Laboratory of Microelectronic Devices and Integrated Technology, Institute of Microelectronics, Chinese Academy of Sciences, Beijing 100029, China}
\affil[6]{University of Chinese Academy of Sciences, Beijing 100049, China}
\affil[7]{Frontier Institute of Chip and System, Fudan University, Shanghai 200433, China}
\affil[8]{Department of Electronic and Computer Engineering, the Hong Kong University of Science and Technology, Hong Kong, China}

\affil[*]{e-mail: zrwang@eee.hku.hk; shangdashan@ime.ac.cn; qi\_liu@fudan.edu.cn}

\begin{abstract}
Human beings construct perception of space by integrating sparse observations into massively interconnected synapses and neurons, offering a superior parallelism and efficiency. Replicating this capability in AI finds wide applications in medical imaging, AR/VR, and embodied AI, where input data is often sparse and computing resources are limited. 
However, traditional signal reconstruction methods on digital computers face both software and hardware challenges. On the software front, difficulties arise from storage inefficiencies in conventional explicit signal representation. Hardware obstacles include the von Neumann bottleneck, which limits data transfer between the CPU and memory, and the limitations of CMOS circuits in supporting parallel processing.
We propose a systematic approach with software-hardware co-optimizations for signal reconstruction from sparse inputs. Software-wise, we employ neural field to implicitly represent signals via neural networks, which is further compressed using low-rank decomposition and structured pruning. Hardware-wise, we design a resistive memory-based computing-in-memory (CIM) platform, featuring a Gaussian Encoder (GE) and an MLP Processing Engine (PE). The GE harnesses the intrinsic stochasticity of resistive memory for efficient input encoding, while the PE achieves precise weight mapping through a Hardware-Aware Quantization (HAQ) circuit.
We demonstrate the system's efficacy on a 40nm 256Kb resistive memory-based in-memory computing macro, achieving 31.5\(\times\), 35.5\(\times\), and 47.2\(\times\) energy efficiency improvements and 10.8\(\times\), 38.8\(\times\), and 6.2\(\times\) parallelism improvements without compromising reconstruction quality in tasks like 3D CT sparse reconstruction, novel view synthesis, and novel view synthesis for dynamic scenes. 
This work advances the AI-driven signal restoration technology and paves the way for future efficient and robust medical AI and 3D vision applications.
\end{abstract}

\begin{document}

\flushbottom
\maketitle

\thispagestyle{empty}
\section*{Introduction}
\begin{figure}[!t]
\centering
\includegraphics[width=0.9\linewidth]{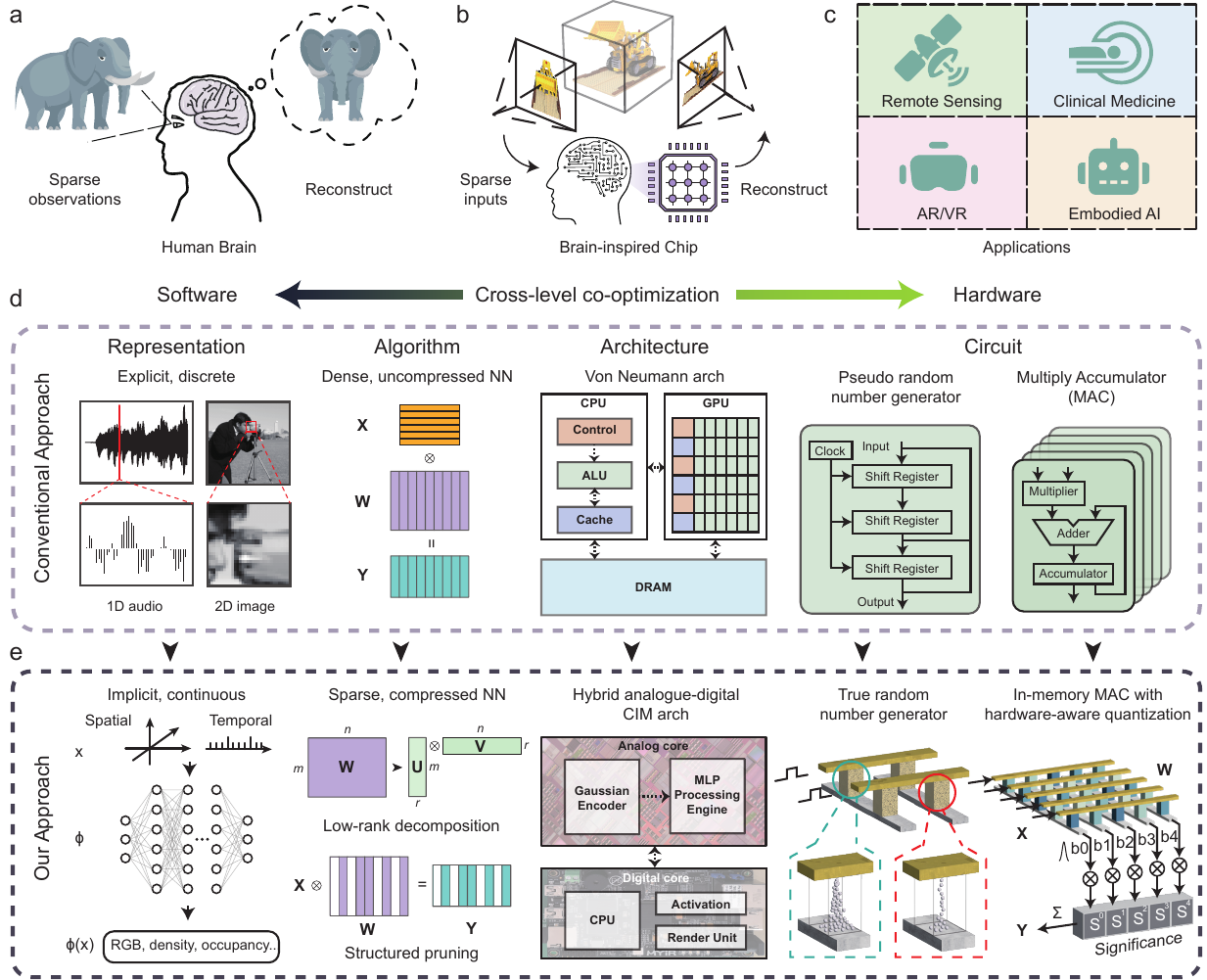}
\caption{\textbf{Cross-level co-optimizations for our system.} 
    \textbf{a,} Human brain's capability of reconstructing experiences from sparse observations.
    \textbf{b,} Our brain-inspired system for efficient signal reconstruction with sparse inputs.
    \textbf{c,} Real-world applications of signal reconstruction from sparse input.
    \textbf{d,} Challenges faced by traditional approaches at different levels of software and hardware. From left to right, these include: At the representation level, traditional explicit representation methods face low storage efficiency, limited flexibility in storage formats, and inadequate scalability for resolution switching. At the algorithm level, uncompressed AI models are unsuitable for edge deployment. At the architecture level, the von Neumann architecture leads to data transfer overhead due to its separate processing and memory units. At the circuit level, frequently used pseudo random number generators and Multiply-Accumulators (MAC) are sequential.
    \textbf{e,} Our approach's innovations across different levels. From left to right, these include: At the representation level, we use neural fields to represent data, with signals as functions of space and time coordinates embodied through a neural network. At the algorithm level, we utilize low-rank decomposition and structured pruning to reduce the number of parameters that need to be mapped onto hardware. At the architecture level, we develop hybrid analog-digital system where the resistive memory-based analog core collocates memory and processing. At the circuit level, we parallelly generate true random numbers using resistive memory's intrinsic randomness for Gaussian encoding, and perform MAC using parallel and precise hardware-aware quantization circuits.
}
\label{fig1}
\end{figure}

The human brain is efficient and fast in constructing perceptions. Humans receive multimodal sparse signal inputs from various senses (e.g., visual, auditory, etc.), implicitly learning representations by tuning plastic synaptic connections of neurons, which allows humans to almost instantaneously reconstruct perceived experiences in their minds at just 20W power\cite{tononi1998complexity} (Fig. \ref{fig1}a). Replicating this ability in AI (Fig. \ref{fig1}b) to efficiently and accurately reconstruct complex signals from sparsely sampled observations finds wide practical applications in remote sensing\cite{shen2015missing}, medical imaging\cite{liu2022recovery}, augmented/virtual reality (AR/VR)\cite{mildenhall2021nerf}, and embodied AI\cite{bartolozzi2022embodied} (Fig. \ref{fig1}c), where the available input data is often sparse and incomplete\cite{santos2023development}.

However, traditional signal reconstruction methods on digital computers encounter multifaceted challenges across both software and hardware dimensions, as illustrated in Fig. \ref{fig1}d.
On the software front, the primary challenge stems from traditional methods of signal representation. Conventionally, signals are represented explicitly by sampling continuous data: audio as discrete-time waveforms\cite{schafer1975digital}, images as pixel grids\cite{rabbani1991digital}, and 3D shapes as voxels\cite{wu20153d}, meshes\cite{karni2000spectral}, or point clouds\cite{qi2017pointnet}. This form of representation requires extensive sampling and storage to achieve high-fidelity reconstruction according to Nyquist–Shannon sampling theorem. 
Consequently, it becomes challenging to maintain high efficiency without compromising on quality.
The second challenge arises at the algorithmic level. In many edge computing applications, such as AR/VR and embodied AI, there is a stringent requirement for real-time processing with limited computational resources\cite{lin2019computation}. Traditional algorithms typically do not employ hardware-aware compression techniques, which leads to high computational burden and energy consumption\cite{han2015deep}, consequently limiting their applications in edge scenarios.
On the hardware side, the third challenge is rooted in the inherent limitations of the von Neumann architecture\cite{horowitz20141}. This architecture is hampered by the von Neumann bottleneck\cite{zidan2018future}, which is characterized by significant overhead in data transfer between the processing and memory units. Such a bottleneck exacerbates energy consumption and limits the speed of signal reconstruction.
Lastly, at the circuitry level, CMOS device scaling is nearing its physical boundaries, slowing down the Moore's Law\cite{wong2015memory}. Additionally, traditional CMOS circuits face limitations in parallel processing\cite{chen2020survey}, especially for heavily used machine learning operations like pseudorandom number generation and matrix multiplication. This bottleneck notably exacerbates system latency, underscoring the difficulties in achieving efficient signal reconstruction within the current software and hardware frameworks.

To address these challenges, we have co-designed a software-hardware framework: neural field for signal reconstruction from sparse input using resistive memory-based computing-in-memory (CIM) hardware (Fig. \ref{fig1}e). 
At the software level, we adopt the implicit neural representation method\cite{sitzmann2020implicit},  encodes information via a function \(f\). This function takes spatial or spatial-temporal coordinates \(\textbf{x}\) as input and outputs corresponding values (such as RGB, density, occupancy, etc.). We parameterize \(f\) using a Multilayer Perceptron (MLP), thereby creating a neural field\cite{hinton2023represent}, capable of approximating complex signals at reduced storage cost compared to explicit methods. 
Secondly, at the algorithmic level, we further reduce the parameter count by employing low-rank (LR) decomposition\cite{jaderberg2014speeding} alongside structured pruning\cite{fang2023depgraph} to train the MLP. LR decomposition involves decomposing each hidden layer \(\textbf{W} \in \mathbb{R}^{m \times n}\) of the MLP into the product of two low-rank matrices, \(\textbf{U} \in \mathbb{R}^{m \times r}\) and \(\textbf{V} \in \mathbb{R}^{r \times n}\) (see Methods for details). The intrinsic rank \(r\) is typically much smaller than the dimensions \(m\) and \(n\), resulting in substantial parameter reduction. Meanwhile, we enhance the network's sparsity through structured pruning, mirroring the sparse neural connections in the human brain\cite{ramsaran2023shift}, and ensuring compatibility with underlying hardware (see Methods for details). The synergy of LR decomposition and structured pruning effectively reduces both the model's parameters and computational load.
At the hardware level, we've developed an emerging resistive memory-based hybrid analogue-digital system. The analog core collocates processing, memory and storage within resistive memory crossbar array to minimize data transfer overhead and enhance energy efficiency \cite{ambrogio2018equivalent,ambrogio2023analog,wan2022compute,wang2017memristors,zhang2023edge,yao2020fully,xia2019memristive,sebastian2020memory,song2017pipelayer,ielmini2018memory,rao2023thousands,yi2023activity,kumar2022dynamical,prezioso2015training,sun2020one,yuan2023neuromorphic}. The digital core complements analogue core in performs operations other than matrix multiplications, such as non-linear activations.
The analog and digital cores form two functional circuit blocks: a random weight Gaussian Encoder (GE) and a high-precision MLP Processing Engine (PE). As resistive memory devices exhibit inherent randomness\cite{cai2020power,wang2023echo,yang2012observation} like that in the biological neural network, we leverage this for input encoding in the GE on the one hand. On the other hand, to realize high-precision matrix multiplication, we developed Hardware-Aware Quantization (HAQ)  that precisely encodes neural network parameters of MLP PE with a novel Variable Current Multiplicative Amplification Circuit (VCMAC). 


In this paper, we demonstrate the effectiveness of our system on a 512\(\times\)512 resistive memory in-memory computing macro using 40 nm technology node. The system exhibits superior performance in several complex tasks such as 3D Computed Tomography (CT) sparse reconstruction, novel view synthesis, and novel view synthesis for dynamic scene. Our co-design achieves 31.5 \(\times\), 35.5 \(\times\), and 47.2 \(\times\) energy efficiency boost and 10.8\(\times\), 38.8\(\times\), and 6.2\(\times\) parallelism boost compared to state-of-the-art GPU while showing 31.68 dB, 26.66 dB, and 29.19 dB average PSNR, respectively, which are comparable to software baselines. 
Our work lays the foundation for future AI-driven signal restoration applications like medical imaging and 3D vision, ultimately bridging the gap between human perception capabilities and AI systems.


\section*{Hardware co-design and system integration}
\subsection*{Hardware-aware quantization}

\begin{figure}[!t]
\centering
\includegraphics[width=0.9\linewidth]{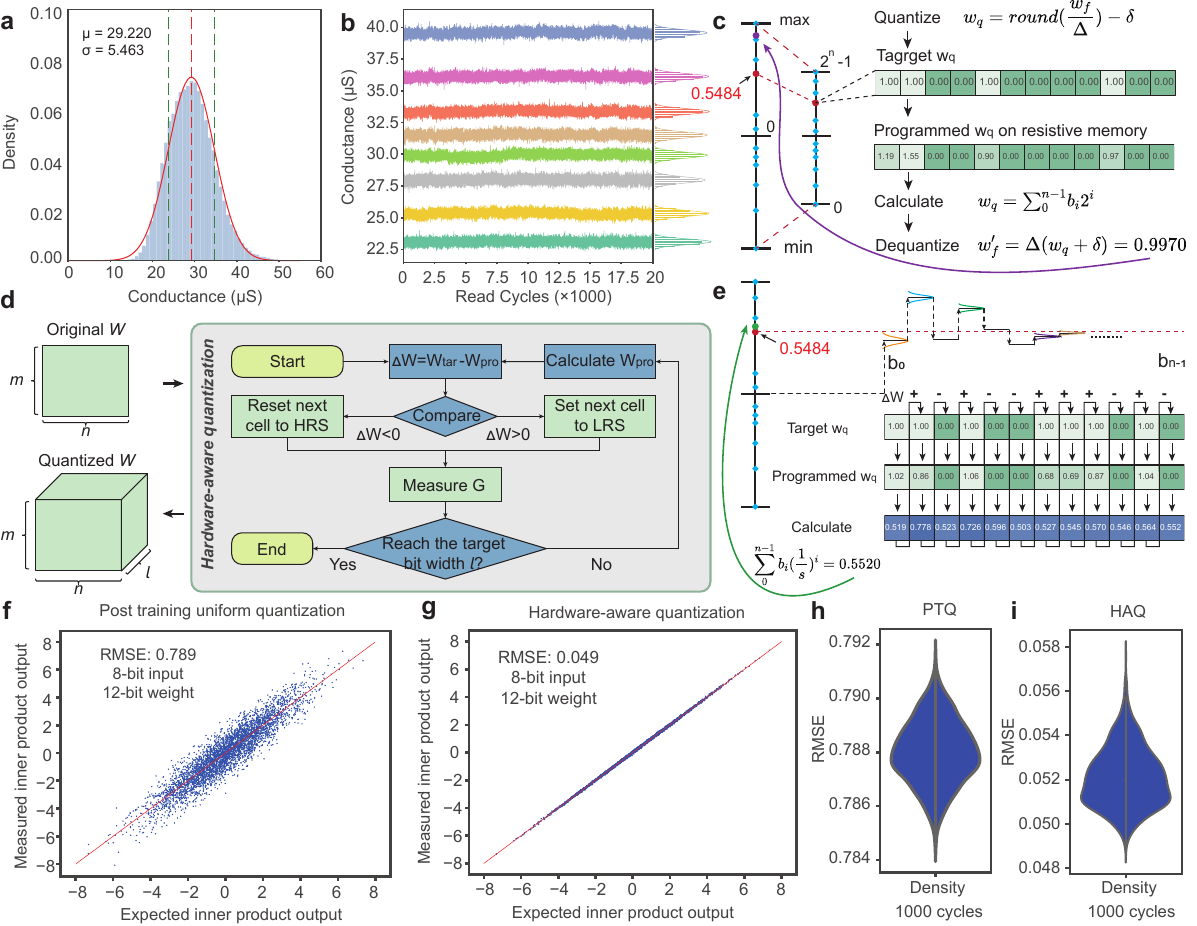}
\caption{\textbf{Hardware-aware quantization (HAQ) for accurate in-memory matrix multiplications.} 
    \textbf{a,} The write noise of resistive memory. The distribution of conductance values read from 10,000 cells subjected to the same set operation.
    \textbf{b,} The read noise of resistive memory. The distribution of conductance values obtained from 8 randomly selected devices during 20,000 read operations.
    \textbf{c,} The quantization process of a weight value using traditional post training asymmetric uniform quantization (PTQ) method. 
    \textbf{d,} The flowchart of the HAQ method. Each cell is iteratively determined to be written as HRS or LRS until the specified bit width is achieved. 
    \textbf{e,} The process of quantizing the same weight value using the HAQ method. The final programmed value closely approximates the original value.
    \textbf{f,g,} The experimental error in matrix multiplication when quantizing with PTQ and HAQ on resistive memory, respectively. HAQ features clear reduction of errors.
    \textbf{h,i,} The stability of matrix multiplication using PTQ and HAQ on resistive memory, respectively. Both methods are robust to temporal conductance fluctuation.
}
\label{fig2}
\end{figure}


Neural field reconstruction has stringent requirements on the weight precision, but resistive memory features inevitable write noise, which stems from the inherent randomness in the electrochemical resistive switching\cite{yang2012observation}. Such randomness can cause deviations between the targeted and the actually programmed conductance in resistive memory (i.e. synaptic weights), leading to errors in the output of machine learning models deployed on these systems. In neural field, as the network output directly represents corresponding field value, which is highly prone to weight errors. 

The hardware noise is revealed by Fig. \ref{fig2}a, the histogram of the conductance values obtained from 10,000 devices upon an identical set programming, exhibiting a Gaussian distribution with a mean of 29.22 \(\mu S\) and a standard deviation of 5.46 \(\mu S\). 
Once programmed, the resistive memory conductance show a decent retention. We performed 20,000 cycles of read operations on these 10,000 devices, with examples shown in Fig. \ref{fig2}b. Even though repeated reading reveals temporal conductance fluctuations, the impact of these read noise on computation is considerably limited compared to the write noise in our system.  

To physically represent model weight using resistive memory, we program resistive memory cells to binary target conductance (low-resistance states, LRS, or high-resistance states, HRS) to present a single bit of a multi-bit weight, to avoid the tedious write-and-verify analogue programming\cite{wan2022compute}.
However, the presence of inevitable write noise yields significant computation errors. 
This is demonstrated in Fig. \ref{fig2}c, the quantization of a weight \(w_{tar}\) using uniform quantization and direct mapping of quantized bits \(W_q\) onto resistive memory. Due to device-to-device variation, each cell does not accurately represent its targeted value, leading to substantial inaccuracies in the actual weights dequantized via \(\sum_0^{n-1} b_i \times 2^i\). This inaccuracy stems from the cumulation of amplified noise across each bit.
Existing quantization method like Post-Training Quantization (PTQ)\cite{banner2019post} or Quantization-Aware Training (QAT)\cite{Jacob2018quant} methodologies all overlook device noise during the quantization process, thereby perpetuating the same issue when deployed directly on resistive memory.

Therefore, we have introduced the HAQ method, where quantization and weight mapping are simultaneous. This method iteratively adjusts for errors introduced by previously quantized bits when determining the next bit, effectively compensating for the cumulative error and significantly reducing the impact of resistive memory write noise. Additionally, this approach can be generalized to various analogue emerging resistive memories. 

Our proposed quantization formula follows \(w=\sum_0^{n-1} b_i\times (\frac{1}{s})^i\). Here, \(b_i\) takes on the values around +1 or -1, achieved by applying a universal bias to the scaled conductance of resistive memory. S represents an adjustable significance ratio of adjacent bits, ensuring HAQ adaptable to different noise levels. Unlike directly programming pre-computed bits \(W_q\) into the resistive memory, we perform quantization bit by bit.
As shown in Fig. \ref{fig2}d, for a given target weight \(w_{tar}\), it is first scaled to the [-1, 1] range. If it's positive (negative), a voltage is applied to set (reset) the most significant resistive memory cell to LRS (HRS), representing +1 (-1). Then, we read the actual conductance value written to the cell and compute its corresponding scaled value. We then calculate the programmed weight \(w_{pro}\) based on the formula and compare it with \(w_{tar}\).
If \(\Delta w\) (i.e., \(w_{pro} - w_{tar}\)) is less (larger) than 0, it indicates that the current weight is smaller (larger) than the target. So we proceed to set (reset) the next cell to +1 (-1) and read its actual conductance to update \(w_{pro}\). This iterative process of comparison, writing, and reading continues until the quantization encompasses the target bit width \(l\), as detailed in Fig. \ref{fig2}e.
Our significance ratio \(s\) is adjustable, allowing us to adapt to various emerging memories of different noise levels and bit width.
Supplementary Fig. 1a illustrates the root mean square error (RMSE) of matrix multiplication employing our HAQ under varying write noises (5\% to 30\%) in simulation for 12-bit quantization. 
It is observed that with increasing noise levels, the optimal \(s\) gravitates towards lower, whereas lower noise scenarios necessitate higher \(s\) values. This trend suggests that the significance ratio \(s\) plays a crucial role in balancing quantization accuracy against inherent noise in the memory system.
We have also conducted simulations to explore the selection of \(s\) for different bit width across 4 to 16 bits, and the results can be found in the Supplementary Fig. 1b, offering further insights into the adaptability of our HAQ method across varying bit resolutions.

We experimentally conducted matrix multiplications on a vector of length 100 using PTQ and HAQ methods on resistive memory for comparison. The inputs were 8-bit, and the matrices deployed on resistive memory were quantized to 12-bit. The matrix multiplication result obtained through the PTQ method yielded an RMSE of 0.789 (Fig. \ref{fig2}f), while the result obtained through the HAQ method yielded an RMSE of 0.049 (Fig. \ref{fig2}g). The HAQ method demonstrated a remarkable 16.1-fold improvement in accuracy compared to the PTQ method, significantly enhancing the precision of matrix multiplication.
Furthermore, in Fig. \ref{fig2}h,i, we illustrate HAQ's resilience to read noise by showcasing the distribution of matrix multiplication RMSE across 1000 reading cycles, which is similar to that of PTQ.
\subsection*{Architecture and circuit}
\begin{figure}[!t]
\centering
\includegraphics[width=0.9\linewidth]{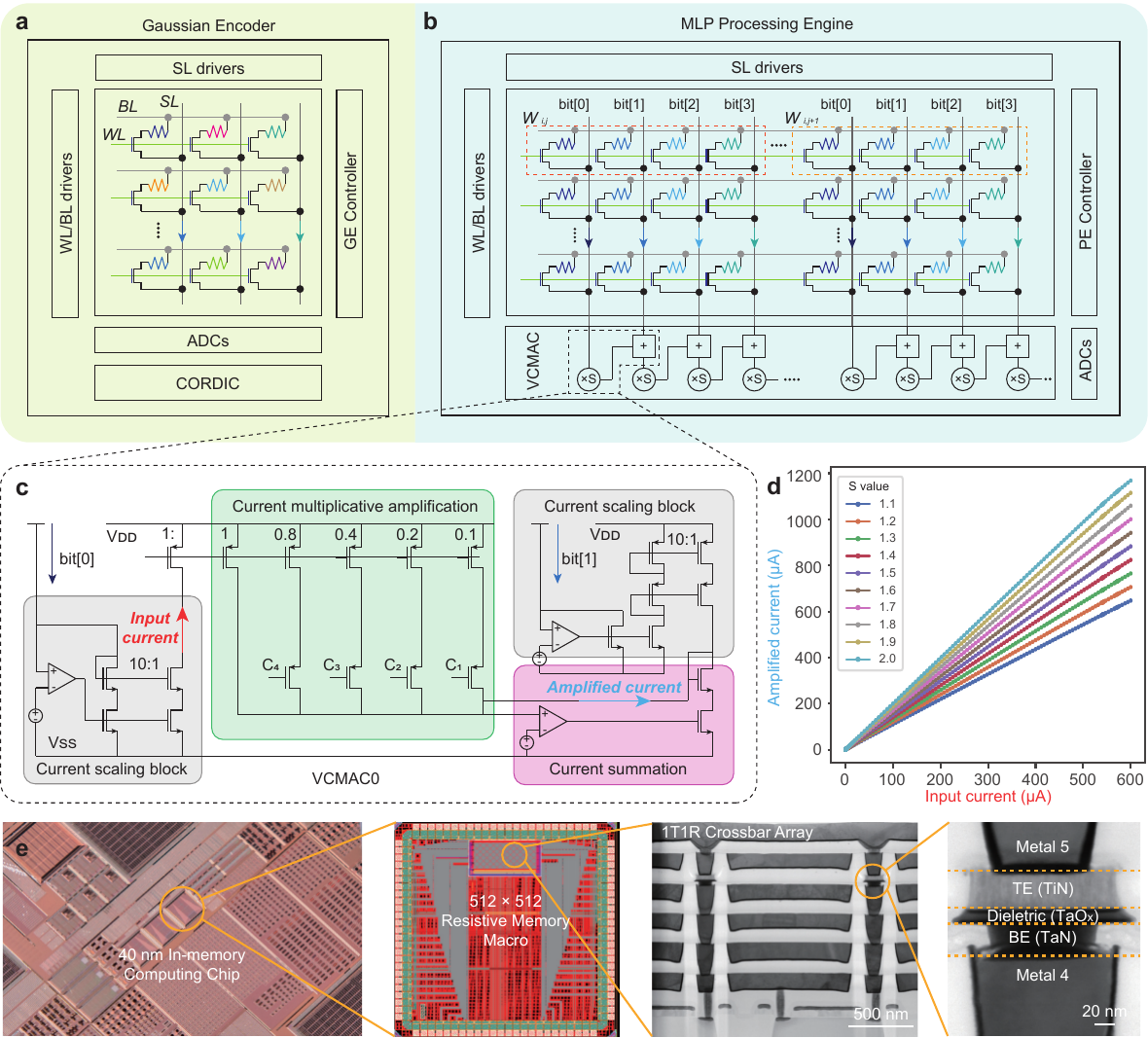}
\caption{\textbf{Architecture and circuits of our hardware co-design.} 
    \textbf{a,} Architecture of the Gaussian Encoder (GE), consisting of a resistive memory in-memory computing block with random weights and a digital CORDIC block.
    \textbf{b,} Architecture of the MLP Processing Engine (PE), consisting of resistive memory in-memory computing block with Variable Current Multiplicative Amplification Circuit (VCMAC) block .
    \textbf{c,} Circuit diagram of the VCMAC block.
    \textbf{d,} Current multiplicative amplification results with VCMAC, where \(S\) represents different significance ratios. The input and amplified current are indicated in Fig. \ref{fig3}.c 
    \textbf{e,} Optical image of the in-memory computing chip, consisting of a 512 \(\times\)512 resistive memory in-memory computing macro using 40 nm technology node, with cross-section transmission electron microscopy (TEM) images of 1T1R array and individual resistive memory cell.
}
\label{fig3}
\end{figure}

Our hardware co-design comprises a Gaussian Encoder (GE) (Fig. \ref{fig3}a) and an MLP Processing Engine (MLP PE) (Fig. \ref{fig3}b), each employs a resistive memory in-memory computing macro as their analog computing core. The difference is that, the Gaussian encoding matrix \textbf{B} leverages the writing noise of resistive memory to produce a stochastic matrix. While the MLP PE precisely maps each weight of the neural field to multiple resistive memory cells using HAQ in the presence of write noise.

In the GE block (Fig. \ref{fig3}a), we augment MLP's learning of complex representations by utilizing Gaussian random encoding. As per Neural Tangent Kernel (NTK) theory\cite{jacot2018neural}, neural networks tend to learn low-frequency information (global features) over high-frequency information (local features). To enable effective learning of high-frequency information, encoding methods like positional encoding map low-dimensional coordinates to different frequency domain spaces, where similar approaches are utilized in the Transformer models\cite{vaswani2017attention}.
Here we develop a hardware-friendly Gaussian random encoding\cite{tancik2020fourier} circuit, where a matrix \textbf{B} sampled from an isotropic Gaussian distribution multiplies the low-dimensional input \textbf{x} and maps it to a richer feature space (see Method for the formula). The random redox reactions and ion migration in resistive memory provide natural randomness (entropy) to physically realize the true random matrix \textbf{B}.
The coordinates are received by the crossbar array via bit lines (BL), multiplied with the random conductance-matrix according to Ohm's Law and Kirchhoff's Law. The output is in the form of current from the source lines (SL), which, after passing through ADCs, proceeds to the digital core for sinusoidal encoding\cite{tancik2020fourier} using CORDIC algorithm\cite{volder1959cordic}.

The MLP PE performs multi-bit matrix multiplication with HAQ encoded weights (Fig. \ref{fig3}b). Weights of a MLP layer are first mapped onto resistive memory using our HAQ method. As mentioned, each resistive memory cell is multiplied by a corresponding significance coefficient \( (\frac{1}{s})^i \), so the value of a composite weight is \(\sum_{i=0}^{n-1} b_i \times (\frac{1}{s})^i\). To efficiently aggregate contributions of different significant bits in matrix multiplication within the analogue domain, we devised a Variable Current Multiplicative Amplification Circuit (VCMAC). Each SL shares the same coefficient, and the current output from each SL is multiplied by this variable coefficient \( s \) and added to the next SL's current to get \( (s \times B_i + B_{i-1}) \). This process continues, amplifying each bit by \( s^{n-1} \) times, resulting in an output equivalent to \(\sum_{i=0}^{n-1} b_i \times (\frac{1}{s})^i\) after scaling the final current output \(\sum_{i=0}^{n-1} b_i \times s^{n-i-1}\) by \((\frac{1}{s})^{n-1}\) in digital domain. The bit width can be flexibly adjusted according to the task.

Fig. \ref{fig3}d illustrates the VCMAC circuit, comprising a current scaling block, a current domain multiplicative amplification circuit, and a current summation component. The current scaling block stabilizes the SL voltage of the resistive memory array, scaling the current by a factor of 0.1 to reduce power consumption. The current domain multiplicative amplification circuit consists of five stages of current mirrors, replicating the scaled current by factors (1, 0.8, 0.4, 0.2, 0.1), controlled by switches \(C_4\)-\(C_1\), allowing for amplification adjustments between 1.1 to 2.5 times. The current summation circuit adds the current from the \(i\)-th bit to the previously computed result of the \(i-1\)-bits.
By switching on and off the current mirrors, we demonstrate current amplification ratios ranging between 1.1 and 2.0, as shown in Fig. \ref{fig3}d). The current amplification is highly precise. The mean amplification error remains within 1\% under normal operating conditions (Supplementary Fig. 2). Furthermore, this error is systematic and can be effectively corrected and eliminated during the aforementioned HAQ process.

The flow of neural field reconstruction is as follows (Supplementary Fig. 3): Input coordinates first enter the GE, where the resultant current output is transmitted to the digital core for periodic encoding. Subsequently, the encoding serves as the input to the model implemented using MLP PE, performing inference using analogue in-memory matrix multiplication with activations provided by the digital core. 

We construct our hybrid analog-digital computing platform based on a resistive memroy in-memory computing macro and a Xilinx ZYNQ system-on-chip on a single printed circuit board (Supplementary Fig. 4). The resistive memory is integrated using a backend-of-line process at the 40-nanometer technology node as shown in Fig. \ref{fig3}e (see Methods), with a crossbar size of 512\(\times\)512. This includes the fabrication of CMOS-compatible nanoscale TaN/TaO\(_x\)/Ta/TiN resistive memory cells with one-transistor-one-resistor (1T1R) configuration, where TaO\(_x\) serves as the resistive switching layer (see Supplementary Fig. 5 for device characteristics). 

\section*{Exerimental results of hardware-software co-design}
\subsection*{3D CT reconstruction}

\begin{figure}[!t]
\centering
\includegraphics[width=0.9\linewidth]{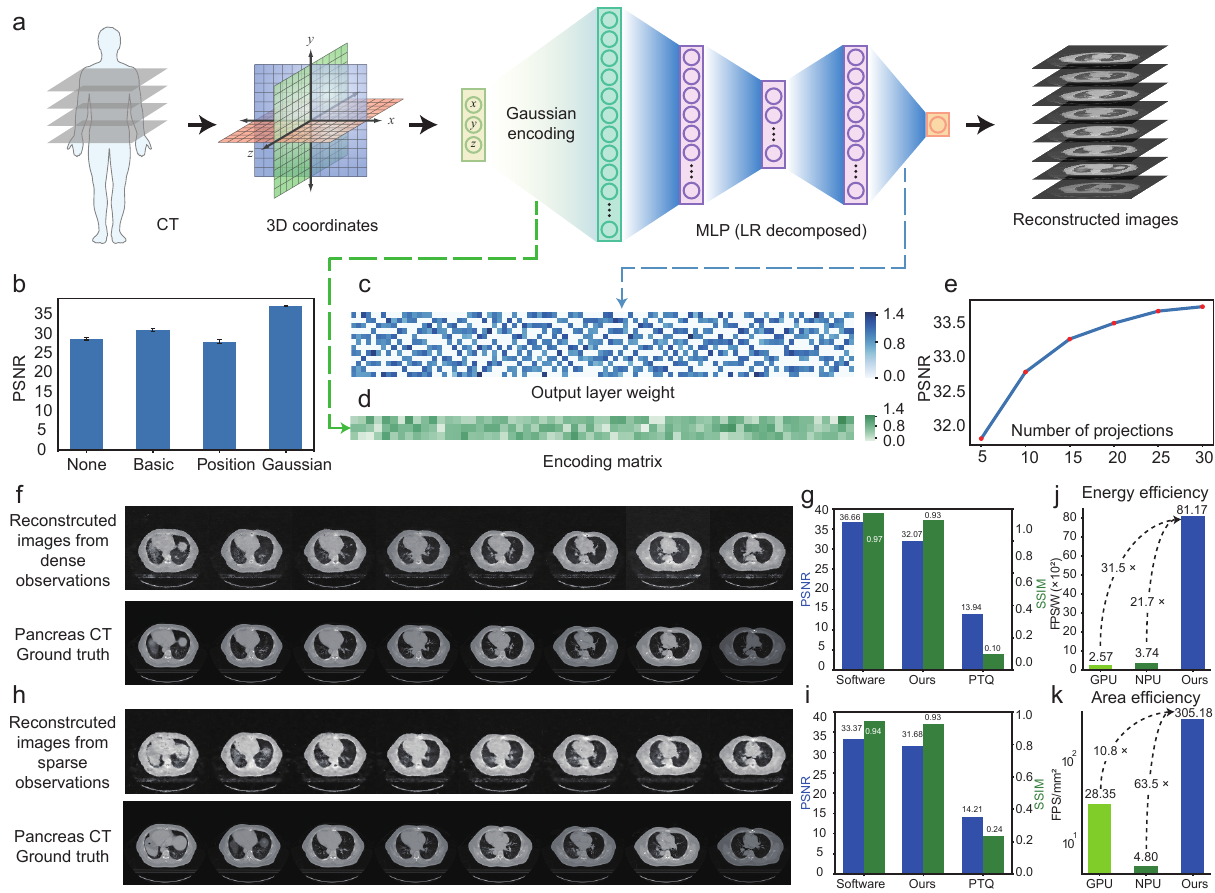}
\caption{\textbf{3D CT reconstruction.} 
    \textbf{a,} Schematic of reconstructing complete 3D CT images from sparse samplings.
    \textbf{b,} The impact of different encoding methods on reconstruction qualities.
    \textbf{c,} The normalized resistive memory conductance distribution of model's output layer weights mapped onto MLP PE through HAQ.
    \textbf{d,} The normalized resistive memory conductance distribution of the random Gaussian encoding matrix of GE.
    \textbf{e,} The reconstruction quality given different number of sparse samplings of CT projections.
    \textbf{f,} The comparison of reconstructed results from dense observations (complete 40 CT slices) and ground truth.
    \textbf{g,} The quantitative comparison of dense reconstruction quality using software, our system with HAQ, and our system with PTQ.
    \textbf{h,} The comparison of reconstructed results from sparse observations (20 CT slices) and ground truth.
    \textbf{i,} The quantitative comparison of sparse reconstruction quality using software, our system with HAQ, and our system with PTQ.
    \textbf{j,} The comparison of CT reconstruction energy efficiency of GPU, NPU, and our system.
    \textbf{k,} The comparison of CT reconstruction area efficiency of GPU, NPU, and our system.
}
\label{fig4}
\end{figure}

In the realm of clinical diagnostics, medical imaging stands as a pivotal tool. Nevertheless, challenges such as radiation dose limits in CT imaging and the slow pace of MRI poses significant hurdles, often leading to under-sampling in the imaging space. This under-sampling constitutes a major bottleneck in achieving high-quality, rapid reconstruction of medical images\cite{chen2008prior, sidky2006accurate}. We demonstrate our system's capability of reconstructing CT images from sparsely sampled data with both low power consumption and high speed.

We evaluated the efficacy of our system in 3D CT image reconstruction through two distinct tasks. The first task involved reconstructing CT fields directly from dense sampling (using all input slices available), demonstrating the ability to reconstruct complex signals. The second task focused on reconstructing CT fields (rebuilding complete CT slices) from sparse inputs, showcasing sparse reconstruction capabilities. For our dataset, we utilized the pancreas 4-D CT data from a clinical patient\cite{shen2022nerp}.

The process of medical image reconstruction is illustrated in Fig. \ref{fig4}a. The 3D coordinates of a pixel \(x\) are given to the GE for encoding, resulting in \(\gamma(x)\). The encoded coordinates are then fed into MLP PE, yielding the intensity of that pixel. Reorganizing these pixels produces the reconstructed CT image.

Fig. \ref{fig4}b illustrates the impact of different encoding methods (see Methods) on the quality of the synthesized images. In this experiment, we maintained the same network architecture while varying only the encoding technique for dense training. A complete CT scan consists of 40 slices, each with \(128 \times 128\) pixels. To ensure a fair comparison, both positional encoding and our random Gaussian encoding were 64 dimensional. It is observed that random Gaussian encoding improved the reconstruction PSNR by approximately 16\%-25\% compared to other methods.

We first use Hardware-Aware hyper-Parameter Optimization (HAPO) to optimize the hyper-parameters of the model on MLP PE (see Methods for details; see Supplementary Fig. 6 for results). Here, the optimal quantization bit length for the input layer, hidden layer, and output layer are 14, 14, and 12, respectively, with a significance ratio of 1.5. We then map the off-line trained weights on MLP PE with HAQ. Fig. \ref{fig4}c and Fig. \ref{fig4}d show the resistive memory conductance map of the output layer on the MLP PE for dense reconstruction, and the random resistive memory conductance map in the GE part of the analog core, respectively. These values, normalized by division through their mean, are distributed between 0 and 1.4.

We further train the model with sparse samplings of CT projection as input, based on the pre-trained model for dense reconstruction. Fig. \ref{fig4}e displays the simulated reconstructed quality as the sparse sampling population varies from 5 to 30 slices. The more input slices used, the better the reconstruction quality. Notably, when the input slice count reaches 20, the model achieves Pareto optimality. Furthermore, it is observed that even with as few as 5 input slices, our model still attains a PSNR of 31.73 dB, indicating a robust reconstruction performance.

Fig. \ref{fig4}f showcases the results of dense reconstruction using our system. Visually, the reconstructed images show no significant differences compared to the ground truth (GT). Fig. \ref{fig4}g quantitatively demonstrates the quality of the dense reconstruction. Compared to software results, hardware reconstruction results with HAQ in terms of PSNR and SSIM are 32.07 dB and 0.93, respectively (for reference, a PSNR over 30 dB or SSIM above 0.9 indicates that the human eye can hardly distinguish between the original and reconstructed images). This represents a decrease of only 12.5\% and 4.1\% compared to software baseline. In contrast, direct hardware reconstruction with PTQ yielded PSNR and SSIM values of 13.94 dB and 0.10, a significant decline of 62.0\% and 89.7\% compared to software, rendering the quality unacceptable for medical diagnosis.
Fig. \ref{fig4}h demonstrates the reconstruction results using our system after learning from sparse inputs, specifically showcasing the reconstruction from 20 CT slices. Eight slices are presented here, and compared to the ground truth, they well retain the anatomical structure and fine details. Fig. \ref{fig4}i quantitatively assesses the quality of sparse reconstruction. Compared to software-based reconstruction, hardware reconstruction results with HAQ for PSNR and SSIM are 31.68 dB and 0.93, respectively, showing only a 5.1\% and 1.0\% decrease. Similarly, direct hardware reconstruction using PTQ resulted in PSNR and SSIM values of 14.21 dB and 0.24, a significant decrease of 57.4\% and 74.5\% compared to software.

We compared the energy and area efficiency of our system in reconstructing CT images. In terms of energy efficiency (fps/W), our system shows 31.5 \(\times\) and 21.7 \(\times\) improvement over a stat-of-the-art GPU (Nvidia A100 GPU) and NPU (Nvidia Jetson Nano) respectively, as detailed in Fig. \ref{fig4}j. Regarding CT reconstruction parallelism, our system also demonstrates 10.8 \(\times\) and 63.5 \(\times\) improvements in area efficiency (fps/mm\(^2\)) compared to GPU and NPU respectively (see Fig. \ref{fig4}k).
\subsection*{Novel view synthesis}
\begin{figure}[!t]
\centering
\includegraphics[width=0.9\linewidth]{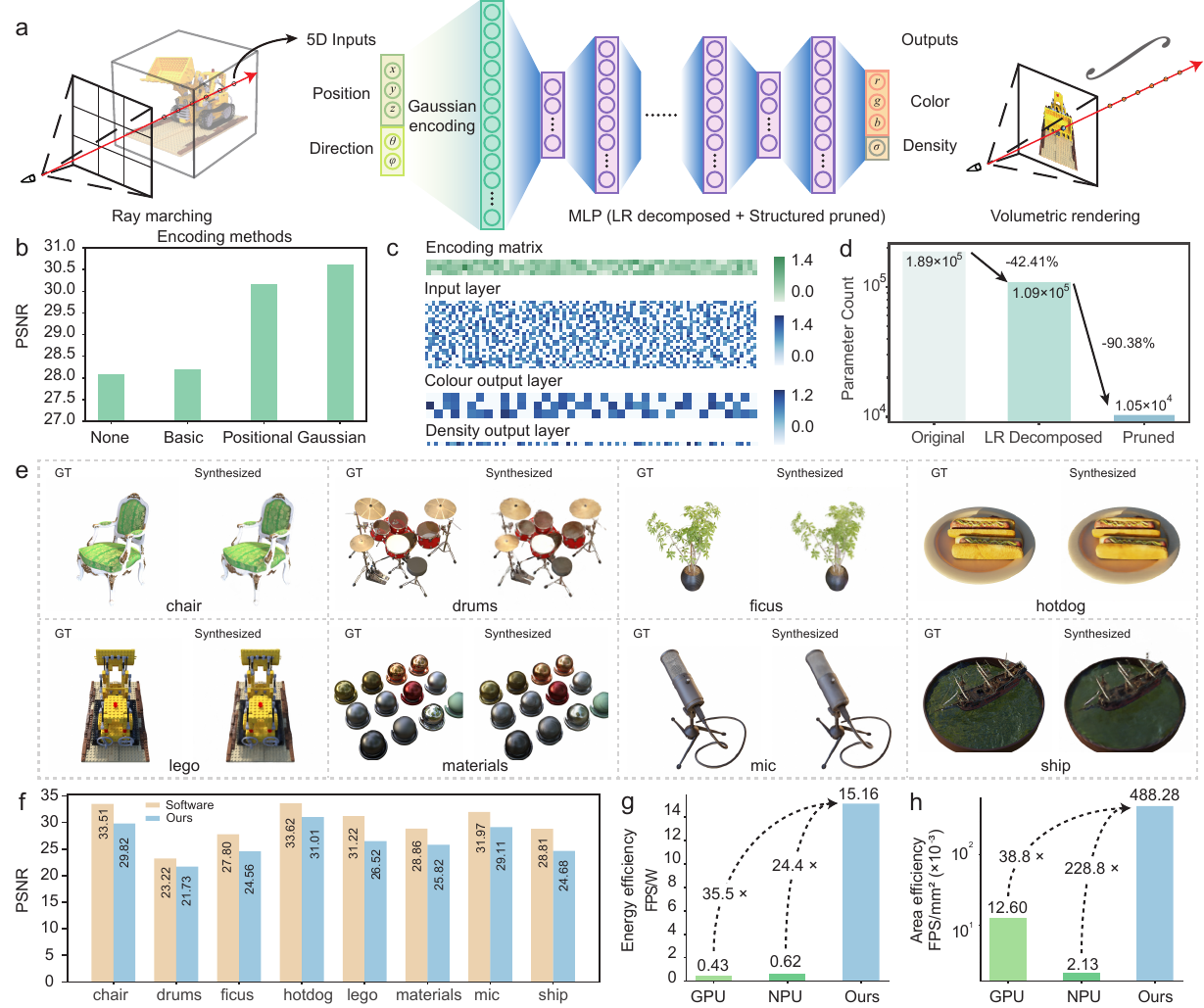}
\caption{\textbf{Novel view synthesis.} 
    \textbf{a,} Schematic of novel view synthesis flow.
    \textbf{b,} The impact of different encoding methods on novel view synthesis qualities.
    \textbf{c,} The resistive memory conductance distribution of random Gaussian encoding matrix in GE, along with the input layer, colour output layer, and density output layer of the model mapped onto MLP PE through HAQ.
    \textbf{d,} The reduction of parameter size using our proposed low-rank decomposition and structured pruning.
    \textbf{e,} The comparison of synthesized novel views using our system, together with ground truth.
    \textbf{f,} The quantitative novel view synthesis results using our co-design, comparable to software baseline.
    \textbf{g,} The comparison of novel view synthesis energy efficiency of GPU, NPU, and our system.
    \textbf{h,} The comparison of novel view synthesis area efficiency of GPU, NPU, and our system.
}
\label{fig5}
\end{figure}

Subsequently, we demonstrate the efficacy of our co-design on the more challenging task of novel view synthesis given a limited input of 2D images using neural radiance field (NeRF). Synthesizing photo-realistic images of 3D scene from novel viewpoints is an important goal in computer graphics as well as generative AI\cite{eslami2018neural}. NeRF-based methods feature promising performance at large inference counts\cite{mildenhall2021nerf}, which benefits from the energy efficiency and parallelism of in-memory computing.

Our co-designed framework is illustrated in Fig. \ref{fig5}a. Our model is a MLP with 8 structured-pruned, low-rank-decomposed hidden layers (see Supplementary Fig. 7 for details). The inputs are spatial coordinates \textbf{x}(\(x, y, z\)) and viewing angles \textbf{d}(\(\theta\), \(\varphi\)), with outputs being color \(\textbf{c}\)(\(r, g, b\)) and density \(\sigma\) of each point. Firstly, a ray is cast from the eye through the pixel of interest, and uniformly sampled along its path. Each sampled point's spatial coordinates and viewing angles are encoded via the GE, before being received by the MLP PE to produce the corresponding color and density. Subsequently, the digital system aggregates these features along the ray through volumetric rendering (see Methods), yielding the RGB value of the given pixel at that viewpoint. Rearranging these pixels forms the complete image.
Fig. \ref{fig5}b presents a comparison of different encoding methods. It is observed that Gaussian encoding effectively improves the final reconstruction qualities by 9.4\%, 8.5\% and 1.7\% respectively, compared to none encoding, basic encoding, and positional encoding. 

The neural network was offline trained before being deployed on our co-design using HAQ. Details on the network's structure, parameters, and training process are available in the Methods section. Hyper-parameters of the co-design were also optimized using the HAPO method. Fig. \ref{fig5}d illustrates the reduction in the parameter count of the compressed model compared to the uncompressed model. Through LR decomposition, the parameter count is reduced by 42.41\%, and further decreased by 90.38\% with structured pruning. Ultimately, this achieves a 18-fold compression without compromising image quality.
Fig. \ref{fig5}c displays the resistive memory conductance distribution of the random GE, as well as the input layer, color output layer, and opacity output layer of MLP PE.

Fig. \ref{fig5}e displays images synthesized from new viewpoints of eight scenes from NeRF synthetic dataset\cite{mildenhall2021nerf} reconstructed by our co-design, each with a resolution of 400\(\times\)400 pixels. Visually, our co-design produced high-quality results. Our system effectively rendered various materials and accurately represented lighting and shadows from different angles, demonstrating its robust rendering capabilities (see Supplementary Fig. 8 for more results).
Fig. \ref{fig5}f quantitatively measures the quality of synthesis of our system with uncompressed software models. On datasets with less complex textures and lighting variations, such as the mic and hotdog datasets, the PSNR values reach around 30. For more complex and detailed datasets like drums and ship, the PSNR is slightly reduced due to limited model size, but still maintains an acceptable visual quality (see Supplementary Fig. 9 for SSIM and LPIPS comparisons). 

We evaluate our system's energy and area efficiency compared to a GPU and an NPU. Fig. \ref{fig5}g shows the energy efficiency (fps/W), where our co-designed system demonstrates a 35.5 \(\times\) and 24.4 \(\times\) improvement over GPU and NPU, respectively. Fig. 5i displays the area efficiency (fps/mm\(^2\)), indicating significant enhancements for the resistive in-memory computing relative to GPU and NPU by 38.8 \(\times\) and 228.8 \(\times\) times, respectively. This performance leap is critical for real-time AR/VR applications.
\subsection*{Dynamic scene novel view synthesis}
We further demonstrate our co-designed system's capability in synthesizing novel views not captured by the original camera setup in dynamic scenes with only limited inputs, a fast-advancing field with wide applications in gaming and cinematography. Dynamic scene synthesis, involving temporal dimension, inherently requires significantly increased computational and storage overheads compared to static scenes. Our hardware offers efficient reconstruction of dynamic scenes, addressing these computational challenges and expanding the scope of real-time rendering applications.

\begin{figure}[!t]
\centering
\includegraphics[width=0.9\linewidth]{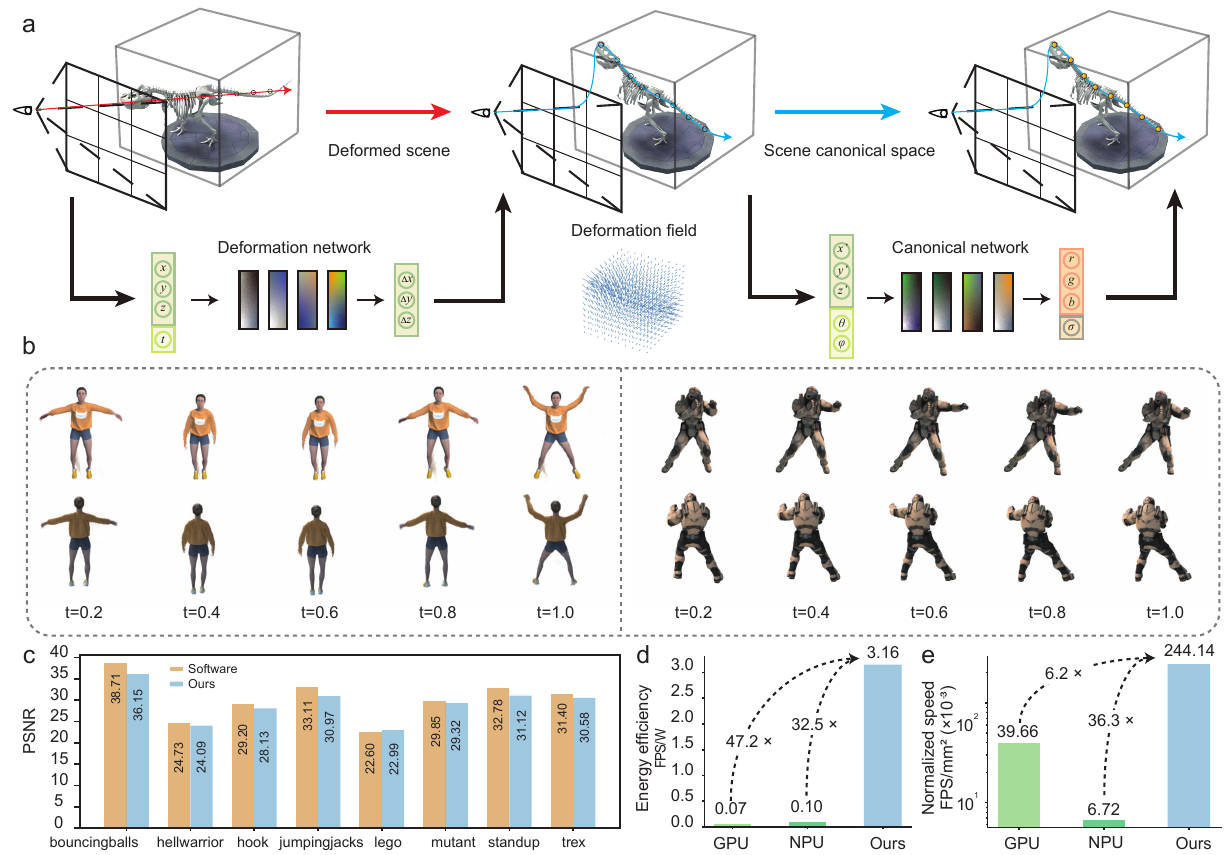}
\caption{\textbf{Dynamic scene novel view synthesis.} 
    \textbf{a,} Schematic of novel view synthesis flow on dynamic scenes. The framework involves two networks, the deformation to produce the object movements, and the canonical network to produce the colour and opacity. 
    \textbf{b,} The comparison of synthesized results from various viewpoints at different timestamps.
    \textbf{c,} The quantitative comparison of dynamic scene novel view synthesis quality using our co-design, the simulated results are comparable to software baseline.
    \textbf{d,} The comparison of dynamic scene novel view synthesis energy efficiency of GPU, NPU, and our system.
    \textbf{e,} The comparison of dynamic scene novel view synthesis area efficiency of GPU, NPU, and our system.
}
\label{fig6}
\end{figure}

Instead of directly incorporating time as an additional input to a single network, we adopted a ray deformation-based method for a better reconstruction quality under the same model size, as shown in Fig. \ref{fig6}a to decouple dynamic and static fields into two smaller neural networks, suitable for CIM hardware deployment (see Supplementary Fig. 10 for details). The first, a deformation network, captures motion and deformation in the scene, taking spatial coordinates and time \textbf{x}(\(x, y, z, t\)) as inputs and outputting the displacement in three directions \(\Delta\)\textbf{x}(\(\delta x, \delta y, \delta z\)). The second, a canonical network, shares the structure with previous NeRF example. It combines the deformation network's output with the current position to get the next moment's coordinates (\textbf{x} + \(\Delta\)\textbf{x}), and, together with viewing angle inputs, produces the final output (color, density). This approach enables the synthesis of new viewpoints in dynamic scenes. 
In our simulation, We first model the distribution of experimental resistive memory programming noise. We then simulate the novel view synthesis of dynamic scences using randomly sampled noisy conductance from the fitted distribution to accurately account for the impact of write noise. Details on the training process are found in the Methods section.

Fig. \ref{fig6}b illustrates the synthetic results over time \textit{t} at different angles on two scenes, Standup and Hook, from the D-NeRF dataset\cite{pumarola2021d}. It is observed that at the given time snap, the synthetic images generated by our system demonstrate high fidelity and fine detail, even in complex scenes involving rigid, jointed, or non-rigid motions, such as human joint movements, etc (see Supplementary Fig. 11 for more synthesized results compared with ground truth).

Fig. \ref{fig6}c quantitatively measures synthesis qualities across eight datasets compared with uncompressed software models. On multiple datasets, the system achieves decent PSNR. For scenes with simpler motion and fewer details, like bouncing balls, the PSNR can reach up to 36. Even in scenes with complex details, such as the Lego and Hellwarrior, the system maintains commendable performance (see Supplementary Fig. 12 for SSIM and LPIPS comparisons).

The benchmark in Fig. \ref{fig6}d,e compares the energy and area efficiency of dynamic scene novel view synthesis of our co-design compared with GPU and NPU. Our co-design is 47.2 \(\times\) and 32.5 \(\times\) higher in energy efficiency, and 6.16 \(\times\) and 36.32 \(\times\) higher in area effeiciency than GPU and NPU, respectively. 


% ...

\section*{Discussion}
In this work, we have designed a resistive memory-based neural field reconstruction solution from both hardware and software perspectives, aimed at achieving efficient, fast and accurate reconstruction of neural fields from sparse data. 
On the software side, we utilize neural field techniques for signal representation, which is augmented by low-rank decomposition and structured pruning to compress the parameter size.
On the hardware front, we have developed a hybrid analog-digital computing system featuring analog in-memory computing on a 40nm 256Kb resistive memory macro. We employ the stochastic nature of resistive memory for random Gaussian encoding, and have developed the HAQ method along with the accompanying VCMAC circuit to achieve high-precision matrix multiplication.
Based on that, we validated our system on 3D CT reconstruction, novel view synthesis, and dynamic scene novel view synthesis, achieving reconstruction results on par with software while significantly saving energy and improving parallelism. 
The software and hardware methods we propose can be adapted to various emerging memory devices, making them a general solution for future signal restoration applications in medical AI, AR/VR, embodied AI, and other related fields.

\section*{Methods}
\subsection*{Fabrication of resistive memory chips}
The memory chip features a 512\(\times\)512 crossbar array composed of resistive memory cells, integrated on a 40 nm standard logic platform. These cells are constructed with bottom and top electrodes, alongside a transition-metal oxide dielectric layer, positioned between the metal 4 and metal 5 layers in the backend-of-line process. The bottom electrodes' vias, 60 nm in diameter, are created through photolithography and etching, and filled with TaN using physical vapor deposition and chemical mechanical polishing. A 10 nm TaN buffer layer is added over the bottom electrode via, followed by a 5 nm layer of Ta, which is then oxidized in oxygen to form an 8 nm TaOx dielectric layer. The top electrodes are made of 3 nm Ta and 40 nm TiN, applied sequentially through physical vapor deposition. Post-fabrication, the chip undergoes standard logic process metal deposition in the logic backend-of-line. Cells in the same column are connected via top electrodes, and those in the same row through bottom electrodes. The chip is finally post-annealed in a vacuum at 400 °C for 30 minutes.

\subsection*{The hybrid analogue–digital computing platform}
This computing system merges a 40 nm random resistive memory computing-in-memory chip with a Xilinx ZYNQ system-on-chip (SoC), both mounted on a printed circuit board (PCB). It supports 64-way parallel analog signal inputs, produced by an 8-channel, 16-bit digital-to-analog converter (DAC80508 from TEXAS INSTRUMENTS), offering a range of 0 V to 5 V. Signal collection involves converting the convergence current into voltages via trans-impedance amplifiers (OPA4322-Q1, TEXAS INSTRUMENTS) and capturing these with a 14-bit resolution analog-to-digital converter (ADS8324, TEXAS INSTRUMENTS). The system includes both analog and digital conversion capabilities on the same board. For vector-matrix multiplication tasks, a DC voltage is applied to the bit lines of the resistive memory chip through a 4-channel analog multiplexer (CD4051B, TEXAS INSTRUMENTS), combined with an 8-bit shift register (SN74HC595, TEXAS INSTRUMENTS). The current output, representing the multiplication result, is converted back into voltages and sent to the Xilinx SoC for additional processing.

\subsection*{Low-rank decomposition}
In implementing the mow-rank decomposition, we diverge from traditional two-phase training then optimization approaches. Our approach initiates training with a low-rank factorization of a hidden layer's weight matrix, traditionally sized \(m \times n\), into two matrices of dimensions \(m \times r\) and \(r \times n\), where \(r\) is the chosen rank. This method significantly reduces the parameter count since the total number of parameters becomes \(m \times r + r \times n\) as opposed to \(m \times n\). The choice of rank $r$ is adjustable based on the specific task and dataset.

\subsection*{Structured pruning}
Our structured pruning technique contrasts with conventional masking-based approaches by physically eliminating weights. This is achieved by removing entire rows or columns in the weight matrix, effectively reducing the number of parameters required to be mapped on hardware. The process of structured pruning is integrated into the training routine, allowing the network to adapt and recalibrate as its architecture evolves. The process involves evaluating and identifying the least significant neurons based on the magnitude of weights. The pruning rate is adjustable based on the specific task and dataset.

\subsection*{Hadware-aware hyper-parameter optimization (HAPO)}
Our HAPO method provides a systematic hyperparameter tuning framework developed for emerging memory to balance between performance and hardware resource utilization. For our task, at the software level, the core hyperparameters are the pruning ratio and the intrinsic rank of the MLP. At the hardware level, the core hyperparameters are the number of bits each layer is quantized to and the Significance ratio of bit.

Our hyperparameter tuning is divided into two steps. The first step targets the software hyperparameters and employs Population-Based Training (PBT) for coarse-grained hyperparameter tuning to find a suitable combination of software hyperparameters. The second step, based on the optimally found hyperparameters, involves a grid search for hardware hyperparameters to finely tune for the best hyperparameters.

In the context of our HAPO method tailored for NVM deployment, we define the optimization objective function as follows:
\begin{equation}
    \text{minimize } \omega \times \frac{PSNR}{PSNR_{\text{max}}} - (1 - \omega) \times \frac{N_{\text{RRAM}}}{N_{\text{max}}}.
\end{equation}
where $\omega$ represents a weighting coefficient, modifiable based on specific task requirements. $PSNR$ denotes the Peak Signal-to-Noise Ratio for a given solution. $PSNR_{\text{max}}$ is the maximal achievable PSNR for the task, serving as a normalization factor. $N_{\text{RRAM}}$ is the number of resistive memory cells utilized in the solution. $N_{\text{max}}$ signifies the maximum number of resistive memory cells available, set to 250,000 in our case to ensure redundancy for error accommodation, given a chip architecture comprising 512\(\times\)512 cells. The coefficient $\omega$ is adjusted to prioritize either performance (higher $\omega$) or resource efficiency (lower $\omega$) based on the constraints and objectives of the specific deployment scenario.


\subsection*{Encoding methods}
We delineate four distinct encoding strategies for input coordinates in coordinate-based MLPs, denoted by \( \mathbf{x} \):

\begin{itemize}
    \item None: No encoding is applied, and the input coordinates are used directly, i.e., \( \gamma(\mathbf{x}) = \mathbf{x} \).
    \item Basic: \( \gamma(\mathbf{x}) = [\cos(2\pi \mathbf{x}), \sin(2\pi \mathbf{x})]^\top \). This technique effectively circumscribes the input coordinates onto a unit circle.
    \item Positional: \( \gamma(\mathbf{x}) = [..., \cos\left(\frac{2\pi j}{\omega} \mathbf{x}\right), \sin\left(\frac{2\pi j}{\omega} \mathbf{x}\right), ...]^\top \) for \( j = 0, ..., m - 1 \). This strategy applies logarithmically spaced frequencies for each dimension, with the scaling parameter \( \omega \) determined through a hyperparameter optimization process. 
    \item Gaussian: \( \gamma(\mathbf{x}) = [\cos(2\pi B\mathbf{x}), \sin(2\pi B\mathbf{x})]^\top \), wherein each element of matrix \( B \in \mathbb{R}^{m \times d} \) is drawn from a Gaussian distribution \( N(0, \sigma^2) \). Here, \( \sigma \) is ascertained for each specific task. 
    In our system, Gaussian encoding is materialized directly through the random forming process of resistive memory, with each element of \(B\) effectively instantiated by the stochastic nature of resistive memory formation. 
\end{itemize}

\subsection*{Benchmarks of image quality}
We employ three widely accepted metrics -- Peak Signal to Noise Ratio (PSNR)\cite{hore2010image}, Structural Similarity Index Measure (SSIM)\cite{hore2010image}, and Learned Perceptual Image Patch Similarity (LPIPS)\cite{zhang2018unreasonable} -- to evaluate the fidelity of reconstructed images. 

PSNR is used to compare the level of distortion between the original and reconstructed images and is defined in relation to the Mean Squared Error (MSE). PSNR is measured in decibels (dB), with higher values indicating better image quality. Generally, an image with a PSNR greater than 30 dB is considered to have good quality. Given an ideal \( m \times n \) monochrome image \( I \) and a reconstructed image \( K \), the mathematical expressions for MSE and PSNR (in dB) are as follows:
\begin{equation}
\mathrm{MSE}=\frac{1}{m n} \sum_{i=0}^{m-1} \sum_{j=0}^{n-1}[\mathbf{I}(i, j)-\mathbf{K}(i, j)]^2
\end{equation}
\begin{equation}
\operatorname{PSNR}=10 \cdot \log _{10}\left(\frac{MAX^2}{MSE}\right)
\end{equation}
where \(MAX\) represents the maximum possible pixel value of the image.

SSIM takes into account factors such as luminance, contrast, and structure. SSIM values range from 0 to 1, with values closer to 1 indicating higher similarity and better quality. The computational formula for SSIM is as follows:
\begin{equation}
    \text{SSIM}(x, y) = l(x, y) \cdot c(x, y) \cdot s(x, y)
\end{equation}
where \( x \) and \( y \) are the original and reconstructed images, respectively. Here: 

\begin{itemize}
    \item Luminance similarity (l):  \(l(x, y) = \frac{2\mu_x\mu_y + C_1}{\mu_x^2 + \mu_y^2 + C_1}\). 
    Here, \( \mu_x \) and \( \mu_y \) are the average pixel values of \( x \) and \( y \), and \( C_1 \) is a small constant.

    \item Contrast similarity (c): \(c(x, y) = \frac{2\sigma_x\sigma_y + C_2}{\sigma_x^2 + \sigma_y^2 + C_2} \). 
    \( \sigma_x \) and \( \sigma_y \) are the standard deviations of pixel values in \( x \) and \( y \), and \( C_2 \) is a small constant.

    \item Structure similarity (s): \(s(x, y) = \frac{\sigma_{xy} + C_3}{\sigma_x\sigma_y + C_3}\). 
    \( \sigma_{xy} \) is the covariance of \( x \) and \( y \), and \( C_3 \) is a small constant.

\end{itemize}

LPIPS is a metric used to measure the perceptual similarity between two images by comparing learned convolutional features. LPIPS scores typically range from 0 to 1, with lower scores indicating higher perceptual similarity and better image quality. The expression is formally defined as:

\begin{equation}
    \text{LPIPS}(x, y) = \sum_{l} \frac{1}{H_lW_l} \sum_{h,w} w_l \cdot \| \phi_l^w(x) - \phi_l^w(y) \|^2 
\end{equation}
where \( \phi_l^w(x) \) and \( \phi_l^w(y) \) are the feature maps at pixel width \( w \), pixel height \( h \), and layer \( l \) for the reference image \( x \) and the assessed image \( y \), respectively. \( H_l \) and \( W_l \) denote the height and width of the feature maps at the corresponding layer \( l \). \( w_l \) represents the learned weights for layer \( l \), which are determined through training on a dataset with human-annotated perceptual differences.


\subsection*{3D CT reconstruction experiments}
The first dense reconstruction task was trained on 40 slices with a resolution of 128 \(\times\) 128. The training process was completed on a dual NVIDIA GeForce RTX 4090 computer.
The second sparse reconstruction task was fine-tuned on the model trained in the first task, and it was trained on 20 slices with a resolution of 128 \(\times\) 28.

The network is a simple MLP with a LR decomposed hidden layer with SIREN activation\cite{sitzmann2020implicit}, a rank of 10, and a hidden layer width of 100. Input coordinates are mapped to 64 dimensions through a random matrix, and after undergoing periodic encoding (sine, cosine), they are concatenated with the unencoded original input to form a vector of length 131, which serves as the network's input. Therefore, the width of the network's input layer is 131, and the output layer width is 1, with the output being the intensity of that pixel point. 
The loss function used is the Mean Squared Error (MSE) loss, and the optimizer is Adam. The learning rate for dense reconstruction was set at \(1 \times 10^{-4}\), and the model was trained for 20,000 epochs. For sparse reconstruction, the learning rate was set at \(1 \times 10^{-5}\), and it was trained for 2,000 epochs.

\subsection*{Volumetric rendering}
We use the volume rendering\cite{mildenhall2021nerf} method to render the color of any ray passing through the scene. The core formula of volume rendering is:

\begin{equation}
C(\mathbf{r})=\int_{t_n}^{t_f} T(t) \sigma(\mathbf{r}(t)) \mathbf{c}(\mathbf{r}(t), \mathbf{d}) d t \text {, where } T(t)=\exp \left(-\int_{t_n}^t \sigma(\mathbf{r}(s)) d s\right)
\end{equation}

This formula represents the expected color of a camera ray \( C(r) \), which is a function of the ray's path \( r(t) = o + td \) extending from the near end \( t_n \) to the far end \( t_f \). The formula also incorporates \( T(t) \), denoting the cumulative transmittance of the ray from \( t_n \) to \( t \), \( \sigma(r(t)) \) representing the volume density at position \( r(t) \), and \( c(r(t), d) \), which is the color at that point.

To numerically estimate the continuous integral for rendering the color \( C(r) \) of a camera ray \( r(t) = o + td \), we use quadrature with stratified sampling. They partition the integral along the ray into \( N \) evenly spaced segments. Within each segment, they sample a point using a uniform distribution. The integral is then approximated as:
\begin{equation}
\hat{C}(r) = \sum_{i=1}^{N} T_i (1 - \exp(-\sigma_i \delta_i)) c_i \text {, where } T_i = \exp\left( -\sum_{j=1}^{i-1} \sigma_j \delta_j \right)
\end{equation}

Here \( \delta_i = t_{i+1} - t_i \) is the distance between adjacent samples. \( \sigma_i \) and \( c_i \) are the density and color at the \( i \)-th sample.

\subsection*{Novel view synthesis experiments}
Our fully connected network architecture, visualized in Supplementary Fig. 7, features input vectors in green, hidden layers in blue, and output vectors in red. Dimensions are noted within each block. The network consists of fully connected layers: layers with ReLU activation (black arrows), layers without activation (orange arrows), and layers with sigmoid activation (dashed arrows). Vector concatenation is denoted by '+'. Gaussian encoding $\gamma(x)$ for input locations passes through eight low-rank approximated fully connected ReLU layers. A skip connection merges this input with activations from the fifth layer. Another layer outputs non-negative volume density $\sigma$ (ReLU rectified) and a feature vector. This vector, concatenated with the viewing direction's position encoding $\gamma(d)$, undergoes processing through an additional fully connected ReLU layer, which outputs emitted RGB radiance at position x for ray direction d.

The network's hidden layer initially has a width of 256 and a rank of 32; after structured pruning, the neuron count is reduced by 90\%, with the width dropping to 26 and rank to 3.
The training process is completed on a dual NVIDIA GeForce RTX 4090 computer. We adopt a batch size of 4096 rays, each sampled at 64 coordinates, and utilize the Adam optimizer with an initial learning rate of \(5 \times 10^{-4}\), decaying exponentially to \(5 \times 10^{-5}\) during optimization. The default Adam hyperparameters are \( \beta_1 = 0.9, \beta_2 = 0.999, \) and \( \epsilon = 10^{-7} \). Optimization for a single scene generally completes in 200k iterations.

\subsection*{Dynamic scene novel view synthesis experiments}
Both the canonical networks \( \Psi_t \) and the deformation network \( \Upsilon_t \) utilize 4-layer MLPs with ReLU activations, and the former includes a sigmoid output. No nonlinearities are applied to \( c \) and \( \sigma \), nor to \( \Delta x \) in \( \Upsilon_t \).
Initial conditions fix the scene at \( t = 0 \) for consistency. 
\begin{equation}
\Psi_t(\mathbf{x}, t)= \begin{cases}\Delta \mathbf{x}, & \text { if } t \neq 0 \\ 0, & \text { if } t=0\end{cases}
\end{equation}

The model trains on 400 \(\times\) 400 images over 800k iterations, with batches of 4,096 rays, each ray sampled 64 times. The Adam optimizer is used with an initial learning rate of \( 5 \times 10^{-4} \), \( \beta_1 = 0.9 \), \( \beta_2 = 0.999 \), and a learning rate decay of \( 5 \times 10^{-5} \).

\bibliography{reference}

\section*{Data availability}
The  pancreas 4-D CT data\cite{shen2022nerp}, NeRF synthetic dataset\cite{mildenhall2021nerf}, D-NeRF dataset\cite{pumarola2021d} are publicly available. All other measured data are freely available upon reasonable request. 

\section*{Code availability}
The code that supports the plots within this paper and other findings of this study is available at \url{https://github.com/SuperFrankyy/Memristive_Neural_Field}. 

\section*{Acknowledgements}
This research is supported by the National Key R\&D Program of China (Grant No. 2022YFB3608300), the National Natural Science Foundation of China (Grant Nos. 62122004, 62374181, 61888102, 61821091), the Strategic Priority Research Program of the Chinese Academy of Sciences (Grant No. XDB44000000), Beijing Natural Science Foundation (Grant No. Z210006), Hong Kong Research Grant Council (Grant Nos. 27206321, 17205922, 17212923). This research is also partially supported by ACCESS – AI Chip Center for Emerging Smart Systems, sponsored by Innovation and Technology Fund (ITF), Hong Kong SAR.


\section*{Competing interests}
The authors declare no competing interests.

\end{document}